%Paper: astro-ph/9409009
%From: RAN@BGUVMS.BGU.AC.IL
%Date: Sat, 3 Sep 94 9:06 0200

\section{\label{int}Introduction\ }

The different images of an astrophysical source, which are generated by a
gravitational lens of solar or higher mass, do not show interference pattern
(Refsdal 1964). This is due to large differences in the travel time of the
light that produces the images. For lenses of much smaller mass ($\sim
10^{-13}-10^{-16}M_{\tbigodot }$), known as femtolenses, the time delay is
much shorter and interference effects are detectable (Schnieder and
Schmid-Burgk 1986, Deguchi and Watson 1986, Peterson and Falk 1991, Gould
1992 and Stanek {\it et. al.} 1993). There is another case in which relative
phases of gravitationally lensed light are important, namely, the case in
which the emitting source is {\em not thermal} but rather, has some internal
spatial correlation (coherence) in the radiated field.

It was shown by E. Wolf (1987, 1989), using statistical wave theory, that
the spectrum of light emitted by such correlated sources changes in the
process of propagation even through free space. In the present work, we
study this spectral modification effect in the presence of a gravitational
lens. More specifically, we consider a planar source in the presence of a
gravitational lens. Using the theory of optics in curved spacetime (see
Schneider, Ehalers and Falco 1992, hereafter SEF) we derive a generalized
form of the {\em Van Cittert Zernike} theorem (Wolf 1970 and Perina 1985).
The {\em Van Cittert Zernike} theorem consists of a simple relation between
the measured coherence of the radiation and the characteristics of the
stochastic emitting source. Our extended version of this theorem includes
also the effect of source correlation and spacetime curvature. Using this
theorem we study the spectrum modification of light propagating through a
gravitational lens.

The rest of the paper is organized as follows: In section \ref{basic} we
review some of the geometrical optics equations in curved spacetime. In
section \ref{csd} we derive the propagation law for the cross spectral
density of radiation emitted by a planar, partially correlated source. The
generalized form of the {\em Van Cittert Zernike} theorem, mentioned above,
is a direct outcome of this law. The effect of source correlations and
gravitational lensing on the measured spectrum is discussed in section \ref
{lineshift}. In section \ref{summery} we summarize and discuss our results.\

\section{\label{basic}The basic equations}

The Maxwell equations in a static gravitational field may be formally
written in the same form as in flat space but, with effective electric and
magnetic permeabilities which depend on the metric tensor $g_{\mu v}$, i.e.,
$\epsilon =\mu =1/\sqrt{g_{00}}$ (Landau and Lifshitz 1971 and SEF). For a
stationary radiation field, in the absence of free charges and currents, the
behaviors of the radiation in the presence of a static gravitational field
is governed by the Helmholtz equation with an effective refractive index $n=%
\sqrt{\epsilon \mu }=1/g_{00}.$

In the case of a gravitational lens, the size of the radiating source is
much smaller than the source - lens and lens - observer distances. Also, the
radiation wavelength is much shorter than all other relevant length scales
including the characteristic variation length of the effective refractive
index $n\left( \vec r\right) $ (corresponding to space curvature). Thus, the
short wavelength and paraxial approximations are valid. Consequently, the
propagation of the field $\left( V(\vec r_p,z,\omega )\right) $ from the
source plane $\left( z=0\right) $ to the observation plane $z$ is described
by the relation:

\begin{equation}
\label{vfield}V(\vec r_p,z,\omega )=\int G(\vec r_p,\vec r_s,z,\omega
)V(\vec r_s,0,\omega )d^2\vec r_s
\end{equation}
where $G(\vec r_p,\vec r_s,z,\omega )$ is the propagator:

\begin{equation}
\label{Green}G(\vec r_p,\vec r_s,z,\omega )=a(\vec r_p,\vec r_s,z)e^{i\frac
\omega c\theta \left( \vec r_s,\vec r_p,z\right) },
\end{equation}
in which (Schulman 1981)

\begin{equation}
\label{aeq}a(\vec r_p,\vec r_s,z)=\sqrt{\det \left[ \frac{\partial ^2\theta
(\vec r_p,\vec r_s,z)}{\partial \vec r_s\partial \vec r_p}\right] }\left(
\frac \omega {2\pi c}\right)
\end{equation}
and $\vec r_s$ and $\vec r_p\,$ are two-dimensional vectors perpendicular to
the z axis in the source and observer planes, respectively. \thinspace The
integration in equation (\ref{vfield}) is to be carried over the entire
surface of the source. $\theta $ is the geometrical optics phase which
depends on a source point $(\vec r_s,0)$ and on the observation point, $%
(\vec r_p,z$ $)$. Every pair of source - observation points are connected by
a ray trajectory $\vec r^{\prime }(\tau ),\vec k^{\prime }(\tau ).\,$
Naturally, the magnification $M$ is defined as the Jacobian relating the
cross section areas of a bundle of rays at the source and the image planes. $%
M$ may also be related to the geometrical optics phase by (see Appendix 1 as
well as SEF and references therein)
\begin{equation}
\label{magnification1}M=\det \left[ D_s\frac{\partial ^2\theta {\bf (}\vec
r_s{\bf ,}\vec r_p,z)}{\partial \vec r_s\partial \vec r_p}\right] =\frac{%
D_s^2\left| a(\vec r_p,\vec r_s,z)\right| ^2(2\pi c)^2}{\omega ^2}
\end{equation}
For later convenience we denote the tensor in the square brackets of
equation (\ref{magnification1}) by $\overleftrightarrow{M}$, i.e., $%
\overleftrightarrow{M}\equiv D_s\frac{\partial ^2\theta {\bf (}\vec r_s{\bf ,%
}\vec r_p,z)}{\partial \vec r_s\partial \vec r_p}$.

\section{\label{csd}The Cross Spectral Density}

In this section we derive the propagation law for the cross spectral density
of radiation emitted from a planar distribution of partially correlated
sources.

The correlation function of the fields at two observation points $\vec
r_{p1},\vec r_{p2}$ is defined as

\begin{equation}
\label{corobser1}W_p(\vec r_{p1},\vec r_{p2},z,\omega )\equiv <V(\vec
r_{p1},z,\omega )V^{*}(\vec r_{p2},z,\omega >
\end{equation}
where the angled brackets indicate two distinct procedures; one is the
ensemble average and the other is an integration $\left( d\omega ^{\prime
}\right) $ over the frequency range $\omega -\Delta \omega $ to $\omega
+\Delta \omega $, where $\Delta \omega $ is much smaller than the line width.

Consider a statistically stationary current distribution (e.g., the
correlation in frequency is a delta function $\delta (\omega -\omega
^{\prime })$). Using equation (\ref{vfield}) in the above equation we get

\begin{equation}
\label{corobser}
\begin{array}{r}
W_p(\vec r_{p1},\vec r_{p2},z,\omega )=\int d^2\vec r_{s1}d^2\vec
r_{s2}\{W_s(\vec r_{s1},\vec r_{s2},0,\omega )a(\vec r_{s1},\vec
r_{p1},z)a^{*}(\vec r_{s2},\vec r_{p2},z) \\
e^{i\frac \omega c\theta (\vec r_{s1},\vec r_{p1},z,\omega )-i\frac \omega
c\theta (\vec r_{s2},\vec r_{p2},z,\omega )}\}
\end{array}
\end{equation}

In the case of a short correlation length and a quasi homogeneous source it
is useful to define the average and difference variables (see Fig.(1) )

\begin{equation}
\label{rs}\vec r_s=\frac{\vec r_{s1}+\vec r_{s2}}2,
\end{equation}

\begin{equation}
\label{rp}\vec r_p=\frac{\vec r_{p1}+\vec r_{p2}}2,
\end{equation}

\begin{equation}
\label{drs}\Delta \vec r_s=\vec r_{s1}-\,\vec r_{s2}
\end{equation}
and

\begin{equation}
\label{drp}\Delta \vec r_p=\vec r_{p1}-\,\vec r_{p2}.
\end{equation}
For convenience we also introduce the notation $\widetilde{W}_p(\vec
r_p,\Delta \vec r_p,z,\omega )\equiv W_p(\vec r_{p1},\vec r_{p2},z,\omega )$
and $\widetilde{W}_s(\vec r_s,\Delta \vec r_s,0,\omega )\equiv W_s(\vec
r_{s1},\vec r_{s2},0,\omega )$. We Taylor expand the phase and the amplitude
in the integrand of equation (\ref{corobser}) with respect to the variables $%
\Delta \vec r_p$ and $\Delta \vec r_s$ around their origin and with respect
to $\vec r_p$ and $\vec r_s$ around the points $\vec r_p^{*}$ and $\vec
r_s^{*}$ which are chosen so that $\left[ \frac \partial {\partial \vec
r_s}\theta (\vec r_s,\vec r_p,z,\omega )\right] _{\vec r_s=\vec r_s^{*},\vec
r_p=\vec r_p^{*}}=0$ . Keeping only the leading terms and using equations (%
\ref{aeq}) and (\ref{magnification1}) we get:

\begin{equation}
\label{genvancit}
\begin{array}{c}
\widetilde{W}_p(\vec r_p,\Delta \vec r_p,z,\omega )= \\ \frac 1{\left( 2\pi
\right) ^2}\exp \left\{ -i\frac \omega c\vec r_s^{*}\cdot \left( \Delta \vec
r_p\cdot
\frac{\overleftrightarrow{M}\left( \vec r_s^{*},\vec r_p^{*}\right) }{D_s}%
\right) \right\} \\ \int d^2\vec r_sd^2\Delta \vec r_s
\widetilde{W}_s(\vec r_s,\Delta \vec r_s,0,\omega )\det \left[ \frac{%
\overleftrightarrow{M}\left( \vec r_s^{*},\vec r_p^{*}\right) }{D_s}\left(
\frac \omega c\right) \right] \\ \exp \left\{ i\frac \omega c\vec r_s\cdot
\left( \Delta \vec r_p\cdot \frac{\overleftrightarrow{M}\left( \vec
r_s^{*},\vec r_p^{*}\right) }{D_s}\right) +i\frac \omega c\Delta \vec
r_s\cdot \left( (\vec r_p-\vec r_p^{*})\cdot \frac{\overleftrightarrow{M}%
\left( \vec r_s^{*},\vec r_p^{*}\right) }{D_s}\right) \right\}
\end{array}
{}.
\end{equation}
\thinspace

For the case of flat space $(\overleftrightarrow{M}=\overleftrightarrow{I})$
and a delta correlated source of the form:

$$
\widetilde{W}_s(\vec r_s,\Delta \vec r_s,0,\omega )=f(\vec r_s,\omega
))\delta (\Delta \vec r_s)\,,
$$
equation (\ref{genvancit}) is reduced to:

\begin{equation}
\label{free}
\begin{array}{c}
\widetilde{W}_p(\vec r_p,\Delta \vec r_p,z,\omega )= \\ \left( \frac \omega
{2\pi c}\right) ^2\exp \left\{ -i\frac \omega c\vec r_s^{*}\cdot \Delta \vec
r_p\cdot \frac{\overleftrightarrow{I}}{D_s}\right\} \int d^2\vec r_sf(\vec
r_s,\omega )\frac 1{D_s^2}\exp \left\{ i\frac \omega c\vec r_s\cdot \Delta
\vec r_p\cdot \frac{\overleftrightarrow{I}}{D_s}\right\}
\end{array}
\,\,.
\end{equation}

This is the {\em Van Cittert Zernike} theorem (Wolf 1970, Perina 1985) which
says that the cross spectral density is the Fourier transform of the
intensity distribution at the source. Equation (\ref{genvancit}) is a
generalization of this theorem which includes also the effect of
gravitational lensing and source correlation.

Equation (\ref{genvancit}) is the main result of the present work. In the
next section we show how it is used to derive the spectrum modification and
the line shifts.

\section{\label{lineshift}Spectrum Modification}

\subsection{\label{genfor}The General Formula for Spectrum Modification}

In order to demonstrate how gravitational lensing and source correlation
affect the observed cross spectral density let us consider a source of the
form:

\begin{equation}
\label{schell}\widetilde{W}_s(\vec r_s,\Delta \vec r_s,0,\omega )=S_0(\omega
)\exp \left\{ -\vec r_s^2/2\sigma _I^2-\Delta \vec r_s^2/2\sigma _\mu
^2\right\}
\end{equation}
\thinspace where $S_0(\omega )$ is the local spectrum, $\sigma _\mu $ is the
correlation length and $\sigma _I$ is a measure of the source dimension. In
this case we get from equation (\ref{genvancit}):

$$
\begin{array}{c}
\widetilde{W}_p(\vec r_p,\Delta \vec r_p,z,\omega )= \\ \left( \frac 1{2\pi
}\right) ^2\exp \left\{ -i\left( \frac \omega c\right) \vec r_s^{*}\cdot
\left( \Delta \vec r_p\cdot
\frac{\overleftrightarrow{M}\left( \vec r_s^{*},\vec r_p^{*}\right) }{D_s}%
\right) \right\} S_0(\omega )\det \left[ \frac{\overleftrightarrow{M}\left(
\vec r_s^{*},\vec r_p^{*}\right) }{D_s}\left( \frac \omega c\right) \right]
\\ \int d^2\vec r_s\exp \left\{ -\vec r_s^2/2\sigma _I^2+i\left( \frac
\omega c\right) \vec r_s\cdot \left( \Delta \vec r_p\cdot
\frac{\overleftrightarrow{M}\left( \vec r_s^{*},\vec r_p^{*}\right) }{D_s}%
\right) \right\} \\ \int d^2\Delta \vec r_s\exp \left\{ -\Delta \vec
r_s^2/2\sigma _\mu ^2+i\left( \frac \omega c\right) \Delta \vec r_s\cdot
(\vec r_p-\vec r_p^{*})\cdot \frac{\overleftrightarrow{M}\left( \vec
r_s^{*},\vec r_p^{*}\right) }{D_s}\right\}
\end{array}
$$

The spectrum is the value of $\widetilde{W}_p$ for $\Delta \vec r_p=0$.
Normalizing this quantity so that for non correlated source the measured
spectrum will coincide with the local spectrum of the source, we get:

\begin{equation}
\label{specmod}
\begin{array}{c}
S(\omega ,\vec r_p)\equiv
\frac{\left| \widetilde{W}_p(\vec r_p,0,z,\omega )\right| }{\sigma
_I^2\sigma _\mu ^2\det \left[ \frac{\overleftrightarrow{M}\left( \vec
r_s^{*},\vec r_p^{*}\right) }{D_s}\left( \frac \omega c\right) \right] }= \\
=S_0(\omega )\exp \left\{ -\left| \frac \omega c\sigma _\mu (\vec r_p-\vec
r_p^{*})\cdot \frac{\overleftrightarrow{M}\left( \vec r_s^{*},\vec
r_p^{*}\right) }{D_s}\right| ^2/2\right\}
\end{array}
\end{equation}
Equation (\ref{specmod}) shows that, the spectrum is modified upon
propagation from the source to the observer. The modifying factor is
$$
\exp \left\{ -\left| \frac \omega c\sigma _\mu (\vec r_p-\vec r_p^{*})\cdot
\frac{\overleftrightarrow{M}\left( \vec r_s^{*},\vec r_p^{*}\right) }{D_s}%
\right| ^2/2\right\} .
$$

\subsection{\label{lineshifts}Line Shifts}

For simplicity, we consider the case in which $(\vec r_p-\vec r_p^{*})$ is
an eigen vector of $\overleftrightarrow{M}$, with an eigen value $M$, so
that we may use $M\left| \vec r_p-\vec r_p^{*}\right| $ instead of $%
\overleftrightarrow{M}\cdot \left( \vec r_p-\vec r_p^{*}\right) $. From
equation (\ref{specmod}) we see that the spectrum is changed. In such a
symmetric case a Gaussian line of the form

$$
\exp \left[ \frac{\left( \omega -\omega _0\right) ^2}{(\Delta _\omega )^2}%
\right]
$$
is changed to

$$
\exp \left[ \frac{\left( \omega -\omega _0\right) ^2}{(\Delta _\omega )^2}-%
\frac{\sigma _\mu \varphi M}{2c^2}\omega ^2\right] ,
$$
where $\varphi =\left| \vec r_p-\vec r_p^{*}\right| /D_s$. Writing this
again as a Gaussian of the form

$$
\exp \left[ \frac{\left( \omega -\omega _1\right) ^2}{(\Delta _{\omega _1})^2%
}\right] ,
$$
we find that the line shift is:

$$
\omega _1-\omega _0=(\Delta _\omega )^2\frac{\sigma _\mu \varphi M}{2c^2}%
\omega _0.
$$
Note that a very large correlation length seems to indicate a very large
shift, however, the shift can not exceed the original frequency range.
Namely, in frequencies where there were no photons in the original spectrum
there will be no photons in the shifted one. Indeed, by equation (\ref
{specmod}) we see that for values of $\omega $ where $S_0(\omega )=0$, $%
S(\omega ,\vec r_p)$ vanishes as well.

To demonstrate this effect we have chosen the following example: A Gaussian
source $S_0(\omega )=\exp \left\{ -\left( \omega \right) ^2/2K_BT\right\} $,
($K_B$ is the Boltzmann constant) where the temperature used is $T=$ $%
10^6\,\,\,^0$k. A lens with a magnification factor $M=10$, and an observer
located at an angle $\varphi =10^{-3}$ rad. off the lens-source axis.
Figures (2)-(5) illustrate the shifts for the cases of correlation lengths :
$\sigma _\mu =(140,70,35,14)\times 10^2$ \TeXButton{TeX field}{\AA}
respectively. Note that for $\sigma _\mu >70\times 10^2$\TeXButton{TeX field}
{\AA} the line shift is greater than the line width.

\section{\label{summery}Summery and discussion}

A few remarks are in order:

\begin{description}
\item  1) Note that interference effects between different regions in the
source are automatically taken into account in our treatment.

\item  2) In general, the lines may be either red or blue shifted depending
on the specific functional form of the spectrum and the correlation
function. However, for a given source, the change in the spectrum is not
likely to be canceled by double lensing (see equation (\ref{specmod}) and
Fig. (2)-(5)).

\item  3) Note that, in a general lens - source configuration, equation (\ref
{specmod}) shows that the spectrum of different images of the same source,
having different $(\vec r_p-\vec r_p^{*})$ and $\overleftrightarrow{M}\left(
\vec r_s^{*},\vec r_p^{*}\right) $, will be distorted in a different way.
\end{description}

Relatively large correlation length may be found in the radiation generated
by collective effects in plasmas. Two examples are:

\begin{description}
\item  i) Curvature radiation emitted by the electrons moving along the
strong magnetic field lines of a pulsar.

\item  ii) Cyclotron radiation from quasars and galactic jets.
\end{description}

Spatial correlations might also be induced by the geometry of the system. An
example is the system of cosmic masers in planet formation regions. In these
regions the star is surrounded by a thin disk in which one or more masers
occur. Originally, the emitted and amplified radiation is not coherent. In
the special geometry where the transverse dimension $a$ of the maser is much
shorter than the longitudinal dimension $L$ (see Fig. (6)), the radiation is
also collimated. Large spatial coherence is ackuired when in addision the
Fersnel number ($F=a^2/\lambda L$, where $\lambda $ is the radiation
wavelength) is small.

A third possibility is that a thermal source radiates into a medium which
has spatial correlations in its fluctuating refractive index (for example a
plasma with a magnetic field in it). In such cases a secondary source is
formed, emitting spatially correlated radiation (Daniel, Malcolm and Wolf
1990).

In the present work we have derived the mathematical tool needed for the
evaluation of the modification of the spectrum upon propagation in curved
spacetime. We have also demonstrated the effect in a numerical example. More
detailed calculations are needed in order to study the effect in more
realistic systems such as the ones mentioned above.

\section{Acknowledgments}

G. H. thanks E. Wolf, M. W. Kowarz and F. V. J. Daniel for illuminating
discussions. R. Z. thanks M. Birkenshaw for helpful discussions. We both
thank V. Usov for fruit full discussions.

\section{Appendix 1}

In this appendix we briefly review the connection between the geometrical
optics phase and the lens magnification.

The ray trajectories are defined by the equations:
\begin{equation}
\label{dxdtdkdt}\frac{d\vec r^{\prime }}{d\tau }=\vec k^{\prime }\equiv
\frac{\partial \theta }{\partial \vec r^{\prime }}\,;\text{ }\,\frac{d\vec
k^{\prime }}{d\tau }=\vec \nabla \,n^2
\end{equation}
with the boundary conditions: $\vec r^{\prime }(0)=\vec r_s$ , $\vec
r^{\prime }(z)=\vec r_p$. Here $\tau $ plays the role of a parameter along
the ray trajectory. The equation for \thinspace $\theta $ along a trajectory
is:
\begin{equation}
\label{dfhidt}\frac{d\theta }{d\tau }=\frac 12k^{\prime 2}+n^2-1\,;\text{ }\,%
\text{ }\theta (0)=0
\end{equation}

Naturally, the magnification is defined by $M=\det \left[ \frac{\partial
\vec r_I}{\partial \vec r{\bf _s}}\right] $, it may also be related to the
phase $\theta $ in the following way: The angle $\vec \delta $ at $\vec r_p$
(see Fig. (7)) is given by:

$$
\vec \delta =\left( k^x,k^y\right) =\left( \frac \partial {\partial x}\theta
,\frac \partial {\partial y}\theta \right) .
$$
The value of $\vec \delta $ at $\vec r_p$, depends also on $\vec r_s$ and:

$$
\vec r{\bf _I}=D{\bf _s\vec \delta (}\vec r{\bf _s,}\vec r_p).
$$
Thus we get

$$
M=\det \left[ D_s\frac{\partial {\bf \vec \delta (}\vec r{\bf _s,}\vec r_p)}{%
\partial {\bf r}_s}\right] =\det \left[ D_s\frac{\partial ^2\theta {\bf (}%
\vec r_s{\bf ,}\vec r_p)}{\partial \vec r_s\partial \vec r_p}\right] .
$$
which is just equation (\ref{magnification1}) in the text.

\section{Figure Captions}

\begin{itemize}
\item  Figure (1): Geometrical representation of the average and difference
vectors.

\item  Figure (2): Line shift for an initially Gaussian spectrum $S_0(\omega
)=\exp \left\{ -\left( \omega \right) /2K_BT\right\} $ with temperature $%
10^6\,\,\,^0$k, correlation length of $\sigma _\mu =140\times 10^2$%
\TeXButton{TeX field}{\AA}, a lens with a magnification factor $M=10$ and an
observer located at an angle $\varphi =10^{-3}$ rad. off the lens-source
axis.

\item  Figure (3): Same as figure (4) but with correlation length of $\sigma
_\mu =70\times 10^2$ \TeXButton{TeX field}{\AA}.

\item  Figure (4): Same as figure (4) but with correlation length of $\sigma
_\mu =35\times 10^2\,$\TeXButton{TeX field}{\AA}.

\item  Figure (5): Same as figure (4) but with correlation length of $\sigma
_\mu =17\times 10^2\,$\TeXButton{TeX field}{\AA}.

\item  Figure (6): An observer seeing the planet forming disk of radius $L$
edge on - will see the radiation emitted from a maser of radius $a$
amplified and collimated.

\item  Figure (7): The geometry of a gravitational lens.
\end{itemize}

\end{document}